# A Framework for Realtime Online Auctions

Bernhard Rumpe
Guido Wimmel

Software & Systems Engineering
Munich University of Technology
80290 Munich, Germany
Phone: +49 (89) 2 89-2 83 62
Fax: +49 (89) 2 89-2 53 10

```
rumpe | wimmel @in.tum.de
http://www.emporias.net/
```

**Abstract**

Among the existing E-Commerce applications, online auctions are the most influential ones. Their impact on trading in the B2B (business to business) as well as in the B2C (business to consumer) and C2C (consumer to consumer) areas will be inevitable. This article describes the architecture of a web-based realtime online auction system, together with the functional and technical requirements that evolved during the development process and heavily influenced the architecture. From the point of view of this real world case study, ways to minimize the development time and yet ensure a robust and flexible system are explained: combining standard software and self-developed components, reusing code wherever possible, and employing the eXtreme Programming approach and its test concepts.

# 1. INTRODUCTION

Electronic commerce will be the enabling technology for the next industrial revolution. Virtual, Internet-based markets allow completely different forms of trading [Höller et. al., 1998]. Local, therefore sometimes monopolistic markets become global and more competitive. The Internet offers a number of different markets. Sellers may advertise, consumers and industrial purchasers can distribute their demands via the Internet. One of the highly compelling and competitive trading forms is based on online auction systems. Auctioning is among the most efficient and fastest concepts to achieve fair prices and identify the optimal business partner. Several forms of auctions allow purchasers to bid for goods and services. Reverse auctions allow suppliers to bid for supplying contracts. Auction forms range from long-term auctions of approx. 4 weeks to short-term auctions of approx. 1h on invitation basis. Auctions may run simultaneously and depend on each other. Multi-phase auctions admit the bidders to the next round only if they hit a certain target price in the previous round. Multi-round biddings enforce one bid per round from each bidder [Prince, 1999].

In running Internet auctions, it became increasingly apparent that the auctioneer must be an independent instance. Only if such an independent auctioneer exists, both supplier and buyer have enough confidence in the fairness of the auction process. The experiences with other marketplaces have shown in an apparent way that an auction marketplace is not very successful when operated by the buyer or seller themselves.

Furthermore, it became obvious that identifying auctionable goods and materials is not an easy task. Therefore, it is a common way to charge a consultant to define the actual auction set-up, starting with the identification of the demands and possible suppliers. On the other hand, it also turned out that it is rather irrelevant to have large supplier lists at hand, because companies that buy material in industrial sizes usually know the their probable suppliers beforehand. They look at the suppliers' situation and they want certification which proves the suppliers' capability of delivering high quality material in time.

In addition to the great variety of possible auction forms, the development of online-auction systems – other E-commerce applications alike – faces particularly tight and often contradicting requirements: the system has to be easy to use, fast, robust, and secure. Even though the field of E-commerce is highly innovative, to our knowledge (state December 2000) no existing online auction system provides all the functionality characterized above [Glänzer, Schäfers, 2000], [Grebe, Samwer, 2000], [Wahrenberg, 2000]. This paper presents a framework for short-term online auctions running in real-time, supporting a multitude of different auction forms, and explains our development process. After two phases of intensive elaboration, www.emporias.net is now under operation.

In the course of the software development process the requirements were gradually comprehended and considerably evolved. Thus, an approach based on eXtreme Programming elements [Beck, 1999] was used to develop the software. This paper overviews the results of this development. The XP approach was the only process that allowed us to develop the business model and the software in parallel, because XP provides enough flexibility to efficiently react to evolving requirements.

Section 2 summarizes the functional and technical requirements that evolved during the software development process. In Section 3 some central elements of the resulting software architecture are presented. Section 4 summarizes the experiences made during the software engineering process for the particular E-commerce application, now online under www.emporias.net (see Figure 1).

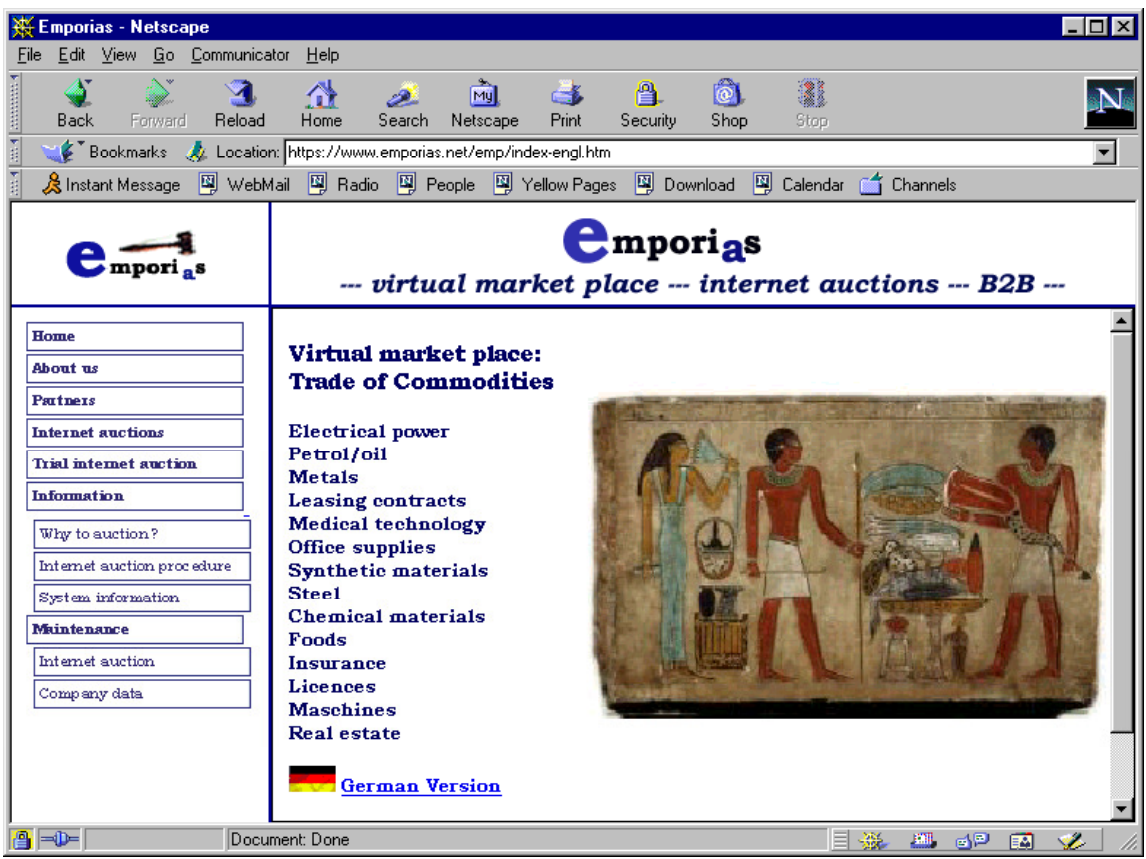

**Figure 1:** Homepage of Emporias.net

## 2. REQUIREMENTS

The product requirements were initially not clear and changed during the project in several major evolutionary steps. This section neither reflects the initial requirements set, nor their history, but a summary of the most recent state of the requirements, which determined the architecture we describe. We distinguish between functional and technical requirements on the software as well as methodical requirements on the software engineering process.

Overall, a lot of common issues encountered in software engineering for web applications can be seen at this example: flexibility, ease of use and a large variety of functionality are usually demanded for on the one hand, but many constraints have to be faced from the technical side, such as security, performance, robustness and compatibility.

*2.1    Functional requirements*

The following list describes the most important functional requirements that evolved during the system development:

1. The software is to be designed for online auctions, both the normal English format and the reverse format, thus allowing to auction goods among buyers and demands among suppliers.

2. An intuitive graphical user interface is to be offered, that must be accessible through the web without any installation necessary.

3. The auctions shall be running in real-time. This means that clients always have current information visible. This is important for short time auctions, where the frequency of bids is relatively high.

4. An auction may consist of several slots, allowing the buyer to split the material desired among several suppliers. This allows to prevent a dependency on a single supplier only, as well as to split delivery.

5. Different auctions may depend on each other. For example, depending on the results of simultaneous auctions, the buyer purchases percentages of competing materials.

6. Persons may participate in an auction in different roles: the auctioneer, the bidders, the originator of the auction (buyer in reverse auctions, seller in the normal auctions), and guests shall be admitted.

7. Different roles get different information at hand. Only the auctioneer can co-relate the bids to their bidders during the auction. Bidders appear to each other anonymously, but know how many competitors there are. Furthermore, bidders see their ranking. External observers following the auction see percentage values instead of real currency.

8. Reverse auctions may have a historic and a target value. The historic value describes what the buyer paid for the auction goods so far, whereas the target value describes what he would like to pay this time. If the auction result hits the target value then the buyer is obliged to sign the contract. If the target prize is not hit the buyer is free to choose.

9. The auction times may vary. Very short auctions may have an auction time as short as 15 minutes. Typical auction times are 1-3 hours, consisting of a main part and an extension part.

10. The auction time is extended whenever a bid arrives shortly before the auction end. This allows all other bidders to react. The provided reaction time may vary, e.g. starting from 3 minutes as an initial extension down to a few seconds at the very end.

11. A login mechanism is imperative. Passwords are distributed through safe channels, among them PGP encrypted emails or SMS.

    12. A report on the auction result is provided for all participants.

For an efficient, correct and robust operation of the auction system, some additional functional requirements emerged as necessary:

1. Both auctioneer and customer shall be able to define auctions themselves. This includes inviting suppliers and defining the access rights and auction format, and shall be handled through the Internet.

2. As auctions are closed (meaning by invitation only), it should be possible to expel participants from a running auction, in case they disregard the rules.

3. For internal administration, a database interface is to be provided (e.g. via the Microsoft Access graphical user interface), and statistical reports are necessary to allow keeping track.

*2.2    Technical requirements*

Whereas in the functional requirements section, requirements and their motivation were discussed, we now give a short summary of technical requirements, and their implications on system architecture:

1. The software shall be downloaded from the Internet without any separate installation being necessary, and should run on currently common hardware and operating systems. Together with the requirement of an active, up-to-date graphic user interface, a Java applet running inside a browser is the best choice.

2. Software shall run even with a low-bandwidth Internet connection, such as a phone line. On the one hand, the downloading time and therefore the size of the applet must be small, which excludes the use of complex frameworks. On the other hand, it is necessary to use a very efficient protocol, excluding Corba or DCOM. Even RMI and XML-RPC proved critical.

3. High security standards are crucial. Any kind of data disclosure or manipulation must be prevented. SSL is the best choice, being supported by common browsers.

4. Privacy. The server is responsible for not revealing any information to any auction participant that the participant is not allowed to see.

5. In short term auctions, the synchronization of client and server times is essential. An appropriate protocol must ensure that the server does not close the auction if a participant still believes it is open. A two-phase dynamic protocol is used for this purpose in our system.

6. The server must be robust with respect to any possible occurring failures.

7. Consistency of the auction data must always be ensured. This makes it necessary to use a robust industrial standard database (Oracle) and define a suitable number of plausibility checks.

8. The realization of the web contents shall be as simple as possible. No extra gadgets, but a full concentration on functionality and user comfort. Of the current technologies for dynamic web pages, the JSP (Java Server Pages) technology supports straightforward development and fits very well into our Java-based architecture.

*2.3    Requirements on Software Engineering Process*

Due to the innovative character of the E-commerce field, several tough requirements for the engineering process apply. For once, E-commerce enforces a high quality of the system. Correctness of the functionality, robustness of the implementation and high security standards are inevitable if money is involved. On the other hand, the current E-commerce development leads to constant evolution and refinement of requirements. The classic software engineering development processes cannot fully cope with instability of customer wishes and state-of-art functionality.

Thus the only appropriate choice was to use a process based on the eXtreme Programming approach [Kent, 1999]. Very short development cycles with weekly releases within the first phase have been very useful. Due to strongly evolving requirements, it is necessary to have an evolution process for the software as well strictly enforcing automatic tests for all important methods and all functionality cases. This is the only way to ensure the high quality of the system.

## 3. ARCHITECTURE

Based on the evolution of the above mentioned requirements during the project a stable architecture evolved. We describe central elements of the architecture of our online auction system, and explain how it fits the requirements discussed in Chapter 2. With our architecture, we developed a well-working compromise for all (partially conflicting) requirements. Furthermore, its concepts can now also successfully be applied to other systems in the field of E-commerce.

*3.1    Overview*

An overview of the system architecture is depicted in Figure 2. It shows one instance per component only, therefore being a conceptual deployment diagram.

The core auction system consists of a number of interacting software components, that together make it possible to conduct, administrate and evaluate online auctions. Like in other web-based E-commerce applications, it is important that the data (auction data, participants, bids etc.) is used consistently throughout the components, that (secure) communication using Internet standards is well supported and that the system is as flexible and platform independent as possible. Therefore, we used Java [Gossling et. al., 2000] as implementation language for most of the system components, allowing for a common data model for online auctions to be shared by these components. For reasons of efficiency and robustness, Apache [Eilebrecht, 1999] was chosen as web server to handle the secure communication with the clients and to dispatch their requests to the respective components behind the firewall. For the same reasons, the industry-standard database Oracle was used to store the permanent data about the online auctions to be conducted by the system. The Java components can easily access this database via the JDBC (Java database connectivity) interface [Hamilton et. al., 1997].

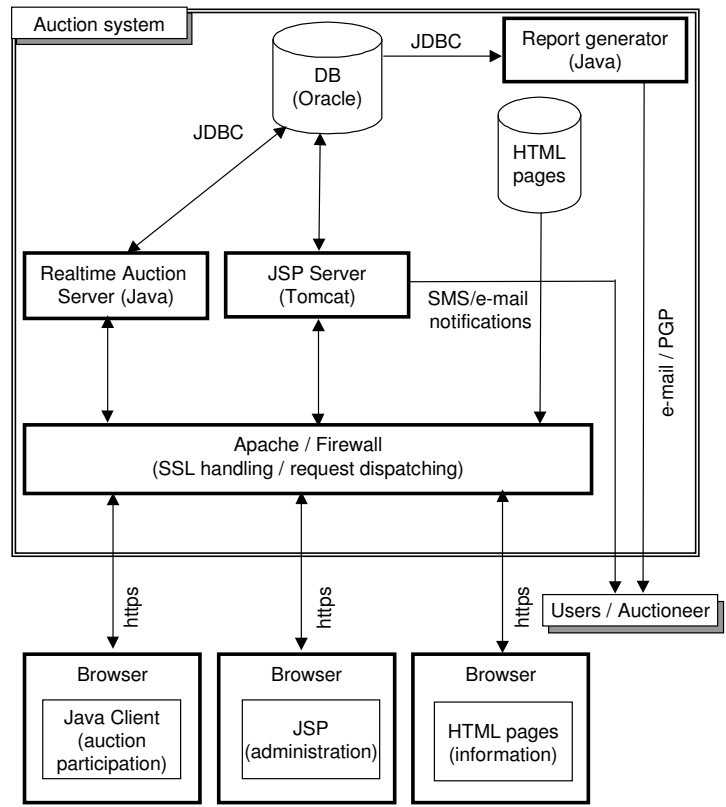

**Figure 2:** Overall system architecture

There are three ways for external interaction with the online auction system:

- A Java client applet, which runs inside the browser while the user participates in the online auctions. The choice of Java, and the fact that all necessary Java classes are downloaded from the web server directly, ensures that the auction system can be accessed from any computer without additional installation effort – all that is required is a browser that supports Java (version 1.1 upwards, e.g. Netscape Navigator $\geq 4.5$ or Internet Explorer $\geq 4.0$).

- JSP pages allow users to view/administrate auction setups. In contrast to other solutions for dynamic web pages like ASP or PHP3, JSP (Java Server Pages) [Fields, Kolb, 2000] can embed Java code. Therefore, the same data-model and infrastructure, i.e. database interface, as for the Java server software can be used.

- HTML pages, simply providing information about the auction system and hyperlinks to the client applet and JSP pages.

Two additional interfaces are for internal use only. They allow to administrate the database and to control and configure the Java processes:

- An auction control interface, which can only be used by the auction administrator and makes it possible to control auctions while they are running (e.g. to ban specific users, to prolong or cancel an auction etc.). The auction control interface was also implemented in Java and communicates with the auction server directly.

- A low-level database administration interface, allowing to edit data in the database or to change the database structure. For reasons of simplicity, Microsoft Access was used, which can interface easily to an Oracle database using the ODBC protocol.

The auction system itself consists of the following components:

- The Apache web server for communication between the other components of the auction system and the users. For security reasons, all communication between the users and the auction system is conducted using SSL. For this purpose, Apache provides a very reliable, fast and robust SSL implementation as a plug-in module. User requests arrive at the Apache Web-Server, where they are directly processed or forwarded to the respective component (i.e. the auction server or JSP-Server). Apache provides the necessary functionality to handle this dispatching, in particular the "mod_jserv" module to forward the requests to the JSP server, and a special kind of "rewrite rule" to forward the requests of the Java client applets to the realtime auction server.

- The realtime auction server is the central part of the system and controls the online auctions. It reads auction data from the database and cooperates with the Java clients to provide them with dynamic information about the auctions, receives their bids and ensures all bids are correctly displayed on the computers of all other auction participants.

- The JSP server, invoked by Apache to handle requests for the dynamic JSP web pages to administrate auction data. The standard JSP implementation Tomcat [Tomcat, 2000] was used for this purpose.

- The report generator, a Java-based program that accesses the database, automatically creates auction reports (e.g. participants, bidding curve, best bid and other statistics) and mails them to the users.

- The Oracle database, storing all the relevant permanent information about the auctions.

The conceptual system architecture given in Figure 2 shows one instance of each process. In order to deal with dynamic load-balancing, processes can be dynamically multiplied on different computers. Running the real auction server behind a number of Apache servers ensures that denial of service attacks are not easy to conduct. An appropriate firewall and traffic monitoring tool supplements the system architecture.

In the following, we will look at some system components in more detail.

*3.2     Oracle Database / Data Model*

All information about the online auctions conducted by the system, as well as current state and given bids are stored in Oracle database. As an industry-standard database, Oracle ensures high reliability, robustness (via logging, recovery and transaction mechanisms), high performance and security (by providing elaborate concepts to protect data from unauthorized access/modification). In addition, robust and well-tested interfaces from the Oracle database to the other components of our system are available through an Oracle implementation of the JDBC interface.

Figure 3 depicts a core part of the data model of the online auction system in form of a UML class model. [OMG, 1999]. Some of these classes like Person, Company and Auction are straightforward. The class diagram also shows that an auction can be split into several slots. Furthermore, several persons of one company may even participate in the same auction having different roles. Only one bidder per company is allowed but several observers may observe what the bidder actually does. The access rights allow a fine-grained definition of which person has which rights to bid or observe which slot at a particular time. Persons may submit as many bids as wanted. Bids are one kind of messages that serves as a communication mechanism between server and client and represent the events that happened during the dynamic execution of the auctions.

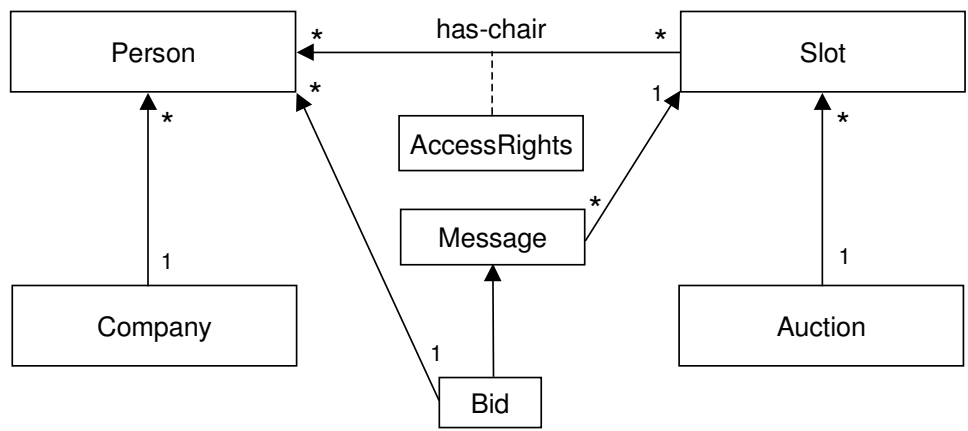

**Figure 3:** Data Model

The data model of Figure 3 is implemented in Java classes as well as the Oracle database. This allows a consistent use of the data model or a subset thereof throughout the system.

In practice this data model is much more complicated. It evolved during the development of the software, which was greatly assisted by the XP Refactoring techniques [Fowler, 1999]

*3.3     Realtime Auction: Client / Server*

The most interesting part of the online auction system is the Java client/server subsystem responsible for carrying out the actual auctions. The client consists of a Java applet and runs inside the Java virtual machine of a browser installed on the user's computer. All information about an auction is stored centrally on the server responsible for that auction - the client merely displays this information and notifies the server if it requests an action (e.g. the user requests a new bid to be placed).

A particularly difficult issue was the time synchronization over the Internet. Due to the rather unstable message delay in the net, a full synchronization is not possible. However, we used a two-phase protocol that ensured the useful average synchronization of client and server time. This includes time differences of the local processor to the server, as well as an overcoming of different time zones in different countries. Only server time is relevant. Experience shows that at the official auction end, the bid ratio is rapidly increasing. Therefore, the auction time is extended up to three minutes to allow each single bidder to react on incoming bids. Still, there is a definitive end, beyond which no bidding is allowed. To ensure that all clients see the auction closed at almost the same time, a two-phase closing protocol is used.

As the client is running in a sandbox, imposed by the browser, there is no chance for the server to directly connect and update the client. Therefore, either the client keeps a channel continuously open or repeatedly asks the server for the new information ("message request"). Unfortunately, most web browsers today are not yet capable of continuously open SSL-secured channels. For almost realtime information, a very short repetition period is necessary (approx. 1 second). This leads to increased traffic. To ensure continuous operation of the system, even with many parallel auctions, the client/server protocol is dynamized, based on a monitor measuring current traffic.

Figure 4 depicts a typical message exchange, denoted by an extended UML sequence diagram. A block in this diagram that essentially depicts a sub-sequence diagram and the middle block shows the repetition and alternative paths.

*3.4    The Basic Client/Server Protocol*

Not to use any additional frameworks at the client side to minimize download time and increase efficiency was an important requirement. The only possibility to implement secure client/server communication via SSL was to use the Java URLConnection mechanism. This is because the implementation of the URLConnection class in the major browsers (Netscape Navigator and Internet Explorer) already contains the necessary SSL functionality - to activate it, the corresponding URL has to begin with "https://").

Therefore, the client/server communication is tunneled over HTTPS. The RPC requests are marshaled by the client and encoded into a HTTPS POST message. Apache carries out the necessary SSL handling and forwards the request to the Java. The server performs the request and sends the result back to the client, again via Apache.

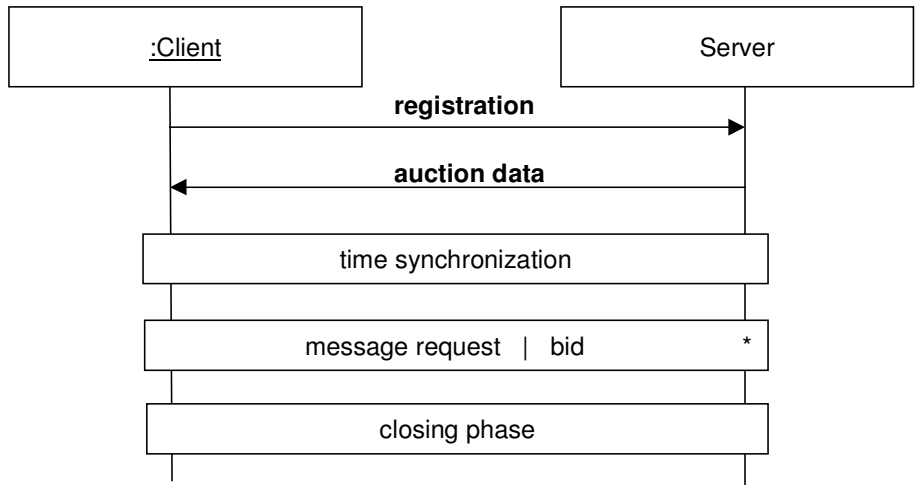

**Figure 4**: Extended Sequence Diagram showing Main Message Exchange

*3.5    JSP*

To allow easy administration, the online auction system offers an interface for customers, as well as consultants to set up auctions themselves. We chose to implement this functionality via JSP (Java Server Pages), as this fits best into our Java-based architecture and the same Java classes modeling the auctions as in the auction server can be used by the JSP pages. According to our auction model, an independent auctioneer ensures a fair auction between the customer and the bidder. This auctioneer gets full access to administrate the auction through the JSP-pages. Behind each auction, there is a workflow model that ensures all auction data a valid, invited bidders are only admitted after they have signed the auction contract, and thereon. The auctioneer can keep track of the status of each potential bidder before the auction starts, e.g. selecting different ways of password transfer.

## 4. SUMMARY OF EXPERIENCES WITH E-COMMERCE ENGINEERING

As it is known, an Internet year lasts only three months. Therefore, new systems have to be built quickly. E-commerce deals with money, in our case with large amounts of money. Therefore, a high-quality standard is inevitable. Rapidly changing requirements, that manifest and change on basis of existing prototypes give a third driving force. These demands do not allow to use the standard software engineering processes with their rather long development phases, and inflexibility. Therefore, we applied a process with a large number of concepts taken from the XP approach [Beck, 1999].

The XP approach goes right to the heart – namely the code. Very short release cycles, no documentation overhead, repeated refocusing on requirements, deep involvement of the customer, and most importantly, the test suite together with existing refactoring techniques characterize the XP approach as a light software development method.

The developed test suites (65% of the code), almost ensure (sic!) that bugs are detected as soon as they are introduced. The refactoring techniques [Fowler, 1999] allow a fast and efficient evolution of the online auction system into its current architecture. This example also shows that is not necessary to define an optimal architecture (in the sense of a class structure) upfront.

The constant application of refactoring allowed us to remove unnecessary code, also, to omit all gimmicks in Html, JSP and Java code. This kept the code clean, small and easy to review.

Both the software and the business model have been developed in parallel. The high flexibility of the used software development approach allowed us to develop both models in parallel and, in particular, to incorporate the new business elements efficiently in the code.

Although we have concentrated on the code, we have used the Unified Modeling Language (UML) to a large extent. Class diagrams showed different portions of the server, client, database and the JSP implementation. Extended versions of sequence diagrams were used to clarify communication protocols. However, we also used the UML to develop tests: concrete scenarios were denoted as object diagrams before they have been mapped to code. Sequence diagrams have been used to describe complex interactions between the code and test-drivers. The mapping of the latter diagrams into test-code was concluded manually, greatly assisted by our own testing framework, which was adapted from JUnit [Beck, Gamma, 2000].

We expect that with appropriate UML-based tools, that allow "constructive generation" of code from UML models as well as generation of "test code" from other parts of the UML models, we would still greatly benefit in engineering E-commerce applications – both in quality and time-to-market. Tools for this purpose are in development.

## Acknowledgements

We would like to thank our colleagues, who helped to develop this software and its business model - in alphabetical order: Samer Al-Hunaty, Julia Bodikova, Manfred Broy, Erich Groher, Andreas Günzler, Robert Heinke, Carsten Jacobi, Klaus Kaluza, Matthias Rahlf, Stefan Schifferer and Horst Wildemann.